\documentclass[conference]{IEEEtran}

\usepackage{mathtools}
\usepackage{amsmath}
\usepackage{amssymb}
\usepackage{caption}
\usepackage[pdftex]{graphicx}
\usepackage{breqn}
\usepackage{epstopdf}
\usepackage{setspace}
\usepackage{bm}
\usepackage{tikz}
\usetikzlibrary{matrix,decorations.pathreplacing}
\newcommand\norm[1]{\left\lVert#1\right\rVert}

\usepackage{algorithm,algorithmic}

\ifCLASSINFOpdf
\else
\fi
\begin{document}
%
\title{A Digital Predistortion Scheme Exploiting Degrees-of-Freedom for Massive MIMO Systems}

\author{\IEEEauthorblockN{Miao Yao, Munawwar Sohul, Randall Nealy, Vuk Marojevic and Jeffrey Reed}
\IEEEauthorblockA{Wireless@VT, Dept. of Electrical and Computer Engineering,\\
Virginia Tech\\
Email: \{miaoyao,mmsohul,rnealy,maroje,reedjh\}@vt.edu}}


%


\maketitle

\begin{abstract}
The primary source of nonlinear distortion in wireless transmitters is the power amplifier (PA). Conventional digital predistortion (DPD) schemes use high-order polynomials to accurately approximate and compensate for the nonlinearity of the PA. This is not practical for scaling to tens or hundreds of PAs in massive multiple-input multiple-output (MIMO) systems. There is more than one candidate precoding matrix in a massive MIMO system because of the excess degrees-of-freedom (DoFs), and each precoding matrix requires a different DPD polynomial order to compensate for the PA nonlinearity. This paper proposes a low-order DPD method achieved by exploiting massive DoFs of next-generation front ends. We propose a novel indirect learning structure which adapts the channel and PA distortion iteratively by cascading adaptive zero forcing precoding and DPD. Our solution uses a 3rd order polynomial to achieve the same performance as the conventional DPD using an 11th order polynomial for a 100$\times$10 massive MIMO configuration. Experimental results show a 70\% reduction in computational complexity, enabling ultra-low latency communications.
\end{abstract}


%
\IEEEpeerreviewmaketitle

\section{Introduction}
\IEEEPARstart{T}{he} power amplifier (PA) is the most power-hungry component in a cellular communications network. The energy efficiency of a base station (BS) significantly depends on the PA efficiency. High input power back-off (IBO) is often required to keep the signal within the amplifier's linear region and avoid in-band and out-of-band distortion and spectral regrowth giving rise to both amplitude/amplitude modulation (AM/AM) and amplitude/phase modulation (AM/PM) nonlinear distortion. However, operating with high IBO results in low energy efficiency for communications and for cooling because of the high heat dissipation. Therefore, the tradeoff between power efficiency and linearity of the PA has motivated the development of linearization techniques for energy-efficient cellular communications. Digital predistortion (DPD), which is used for PA nonlinearity compensation, is a popular solution among the linearization schemes because of its flexibility, accuracy and ability to adapt to the time-variant characteristic of the PA. It enables the application of low-cost PAs and operation of higher signal levels for increased energy efficiency without sacrificing linearity. This is especially needed for the BS of 5G new radio (NR) since it will use one or two orders of magnitude more PAs than a 4G LTE BS. 

The DPD aims to realize a linear response for the combined DPD-PA block by cascading the PA and its inverse response. It can be generally categorized into two groups: polynomial based scheme \cite{mirri2002modified} and look-up table (LUT) based scheme \cite{zhi2006improved}. The Volterra series represents one of the most popular polynomial models of a nonlinear system. Moreover, the PA power response is time-variant when dealing with wideband signal inputs such as 10 or 20 MHz LTE. Therefore, the memory-based Volterra model is widely applied to represent the behavior of a PA in wideband scenarios. Literature shows that a wide range of PA nonlinearities can be approximated with considerable precision by the Volterra series-based filter of sufficient order and memory depth in both single-input single-output \cite{mirri2002modified} and multiple-input multiple-output (MIMO) systems \cite{suryasarman2015comparative,abdelhafiz2016high}. The Volterra series-based method can be a complete representation of an unknown nonlinear dynamic system, but has the drawback of requiring a large number of basis functions \cite{yu2012band,ma2014wideband,bensmida2016extending}. The associated computational complexity limits its practical application especially for large-scale MIMO (a.k.a. massive MIMO) systems.

MIMO technology has gained popularity in the development and deployment of modern wireless communication systems, because of the improved signal to noise ratio (SNR), link coverage and spectral efficiency. To accommodate for the increasing demand of smart devices in emerging economies as well as heterogeneous networks, proposals for 5G NR wireless communications standards consider massive MIMO with up to hundreds of transmit antennas. This means that up to hundreds of PAs, one per antenna are needed. The prohibitive cost of highly linear PAs for massive deployment requires the use of low-cost PAs. In massive MIMO systems, a digital predistorter is needed for each low-cost PA. The implementation of digital predistorters in massive MIMO system is challenging for the following reasons: 
\begin{itemize}
\item	The use of full-order and hence highly precise Volterra series-based DPD is computationally impractical; 
\item	The required number of Volterra series grows exponentially with the number of antenna when we are considering the notable RF/antenna crosstalk in MIMO systems  \cite{suryasarman2015comparative}; and 
\item	The input of the PA is derived from the precoder/beamformer, which combines multiple data streams and hence creates an interplay between dynamic precoder and DPD.
\end{itemize}
\begin{figure*}[h!]
\centering
\includegraphics[width=7.2in]{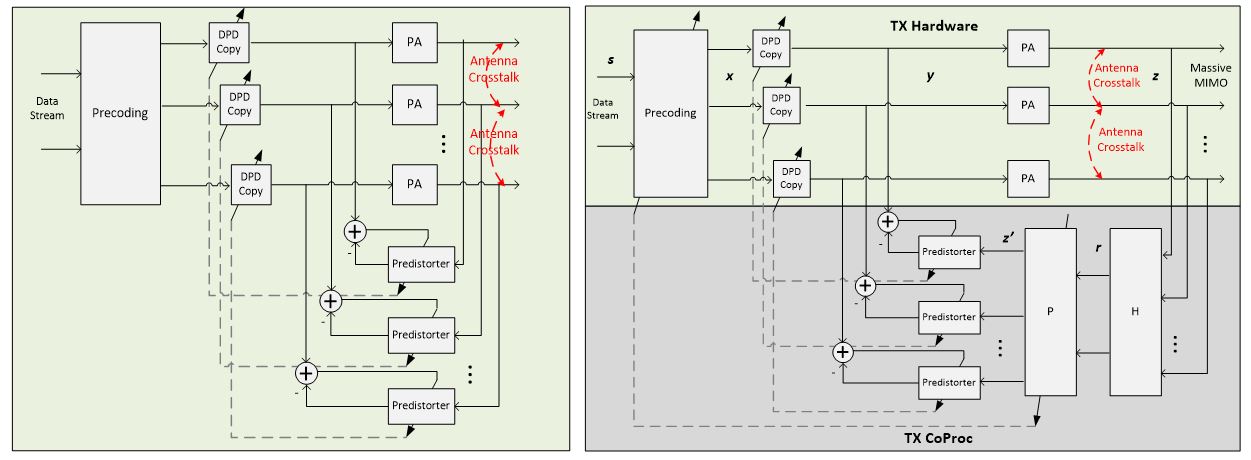}
\caption{Conventional DPD for massive MIMO systems (left) and proposed, low-complexity and precoding-aware DPD solution (right, TX hardware: transmitter hardware, TX CoProc: transmitter coprocessor, the feedback loop includes PAs, channel matrix H which is derived from the TDD uplink channel estimation, adaptive precoding matrix P, and adaptive predistorter).}
\label{fig_block_DPD}
\end{figure*}

Therefore, it is critical to extend the operation of the PA into weakly nonlinear region of the low-cost PAs in massive MIMO systems and hence reduce the required order of the Volterra series \cite{peng2016digital,wood2017system}. This paper proposes a low-complexity and scalable DPD architecture and algorithm that take advantage of the spatial degrees-of-freedom (DoFs) of massive MIMO to facilitate energy-efficient 5G NR BSs. 

The rest of the paper is organized as follows. Section II introduces the system and signal model. Section III presents the
proposed successive refinement filtering structure and computational complexity in terms of hardware implementation. Section IV discusses the numerical and experimental results, and Section V concludes the paper.

\section{System Model and Problem Formulation}
\label{sec:sysmodel}
Consider a downlink massive MIMO system which has one BS equipped with $N_t$ antennas. The number of BS antennas is significantly larger than the number of single antenna users, i.e. $N_t\gg M_r$. When the BS transmits signals to $M_r$ users in parallel by means of spatial multiplexing, the received signal of user $k$ can be represented as
\begin{equation}
r_k=\mathbf{h}_k\Big[f_1(\mathbf{p}_1\mathbf{s}),f_2(\mathbf{p}_2\mathbf{s}),\cdots,f_{N_t}(\mathbf{p}_{N_t}\mathbf{s})\Big]^T+n_k,
\end{equation}
where $\mathbf{h}_k$ is the $k$th row of the channel matrix and represents the $1\times N_t$ channel vector from the BS to user $k$, $\mathbf{p}_k$ the $1\times M_r$ precoding/beam-weight vector, $\mathbf{s}$ the $M_r\times 1$ information symbols, $f_i(\cdot)$ the nonlinear amplification operation of the $i$th PA, $n_k$ the zero-mean complex circular-symmetric additive white Gaussian noise (AWGN) at user $k$, and $T$ the matrix transpose operation. The information symbols at the BS are mapped to the appropriate transmit antennas so that the information received by each user has minimal interference from the signals of the other users. A larger number of spatial DoFs necessary for zero forcing allows selecting from a larger signal space for precoding to minimize the multiuser interference. It allows the PAs to work near the nonlinear operating region.

In this section, we start with an ideal PA model by assuming that the PAs are perfectly linear and ignoring the memory effect to simplify the analysis. Zero forcing precoding (ZFPC) is employed for multiuser signal transmission because of its simplicity and outstanding performance. The beam-weights of ZFPC need to satisfy the constraint
\begin{equation}
\mathbf{H}\mathbf{P}=\mathbf{I}_{M_r},\label{zfpc}
\end{equation}
where $\mathbf{H}=[\mathbf{h}_1^T,\cdots,\mathbf{h}_{M_r}^T]^T$ denotes the $M_r\times N_t$ channel matrix between the BS and all users, $\mathbf{P}=[\mathbf{p}_1^T,\cdots,\mathbf{p}_{N_t}^T]^T$ the $N_t\times M_r$ precoding matrix and $\mathbf{I}_{M_r}$ the $M_r\times M_r$ identity matrix.

Without loss of generality, the PA nonlinearity and the RF coupling between the different paths of the transmitters, which are the two major sources of ZFPC condition violation and the transmitter impairment, need to be considered for massive MIMO deployments. Therefore, an extra nonlinear function is needed to preprocess the input signal of the PA and thus linearize the overall cascaded DPD-PA amplified signal. In order to determine this nonlinear function, the behavioral model and inverse behavioral model of PAs are needed.

Suppose that $f_i(x_i)$ is the nonlinear model of the transmitter for the baseband input signal $x_i=\mathbf{p}_i\mathbf{s}$ and $z_i$ is the equivalent complex signal at the output of the $i$th PA. The predistortion function $g_i(x_i)$ then has to satisfy the following set of relations
\begin{equation}
z_i=f_i(g_i(x_i))=G_0 x_i,
\end{equation}
where $G_0$ is the equivalent linear gain of the DPD-PA cascaded system. A Volterra series-based memory polynomial predistorter is applied to compensate for the dynamic nonlinear behavior. This model, known as the parallel Hammerstein model in literature \cite{suryasarman2015comparative}, is a parallelization of a nonlinear function followed by a linear memory. Moreover, apart from the typical odd-order polynomial representation, even-order polynomials are also included to enrich the basis set and improve the modeling accuracy and DPD performance \cite{suryasarman2015comparative}. The discrete baseband-equivalent form of the Volterra series with memory effect consists of a sum of multidimensional convolutions that can be written as \cite{suryasarman2015comparative}
\begin{equation}
y(n)=\sum_{q=0}^Q\sum_{k=1}^K\omega_{k,q}|{x(n-q)}|^{k-1}x(n-q),\label{y_n}
\end{equation}
where $x(n)$ and $y(n)$ are the input and output complex envelopes, $\omega_{k,q}$ the polynomial coefficient of the filter tap for the $k$th order and $q$th delay, $K$ the nonlinear degree, and $Q$ the memory depth. By defining a new sequence,
\begin{equation}
r_{k,q}(n)=|{x(n-q)}|^{k-1}x(n-q),
\end{equation}
we can rewrite ($\ref{y_n}$) as
\begin{equation}
\mathbf{y}=\mathbf{R}\bm{\omega},
\end{equation}
where $\mathbf{y}=[y(n),y(n+1),\cdots,y(n+N-1)]^T$, $\bm{\omega}=[\omega_{10},\cdots,\omega_{11},\cdots,\omega_{KQ}]^T$, $\mathbf{R}=[\mathbf{r}_{10},\cdots,\mathbf{r}_{11},\cdots,\mathbf{r}_{KQ}]$, and $\mathbf{r}_{kq}=[r_{kq}(n),r_{kq}(n+1),\cdots,r_{kq}(n+N-1)]^T$. Notice that the proposed Volterra-based models are linear with the coefficients $\bm{\omega}$. We can then estimate the model parameter by minimizing the cost function $\norm{\mathbf{y}-\mathbf{R}\tilde{\bm{\omega}}}^2$ \cite{suryasarman2015comparative}:
\begin{equation}
\bm{\omega}=\text{arg}\min_{\tilde{\bm{\omega}}}\norm{\mathbf{y}-\mathbf{R}\tilde{\bm{\omega}}}^2.
\end{equation}
In this paper, we apply the least mean squares (LMS) adaptive algorithm for DPD, where the DPD coefficients are updated every sample. The LMS algorithm is based on the minimum mean square error rule with steepest descent which can be described as \cite{suryasarman2015comparative}
\begin{subequations}
\begin{align}
 \mathbf{e}(m)&=\mathbf{y}(m)-\mathbf{R}(m)\tilde{\bm{\omega}}(m-1),\\
 \tilde{\bm{\omega}}(m)&=\tilde{\bm{\omega}}(m-1)+\mu \mathbf{e}^H(m)\mathbf{R}(m). 
\end{align}
\end{subequations}
The weights of the LMS filter $\bm{\omega}(m)=\text{LMS}(\mathbf{x}(m),\mathbf{y}(m))$ are a function of the input $\mathbf{x}(m)$ and output $\mathbf{y}(m)$ of the LMS algorithm, initialized as $\bm{\omega}(0)=\bm{\omega}_0$. Expression $\mathbf{e}^H(m)$ is the Hermitian transpose of the error vector and $\mu$ the step size.
\section{Proposed Architecture and Algorithm}
The predistorter is trained using the system identification architecture, where the PA characteristics are identified in the feedback path \cite{suryasarman2015comparative}. A conventional indirect learning structure applied for massive MIMO system is shown in Fig. $\ref{fig_block_DPD}$ (left) \cite{liu2015general,zhai2014nonlinear}. It duplicates the DPD-PA structure for each RF path and the LMS filters associated with each PA are updated independently. This architecture fails to consider the influence of ZFPC on DPD.

Since the number of BS antennas in our model is significantly larger than the number of users, the precoding matrix is underdetermined. This implies that there is more than one solution for the precoding matrix satisfying the constraint (\ref{zfpc}). The number of solutions depends on the rank of the channel matrix. The objective of our proposed scheme is finding the appropriate precoding matrix which enables low-complexity DPD with low order basis functions. Therefore, instead of the traditional one-stage indirect learning structure of Fig. $\ref{fig_block_DPD}$ (left), we propose a novel learning structure which adapts the channel and PA distortion iteratively by cascading adaptive ZFPC and DPD. We modify the conventional indirect learning architecture by incorporating the channel matrix into the feedback path. To be more specific, the feedback loop in Fig. $\ref{fig_block_DPD}$ (right) includes the channel matrix, adaptive ZFPC, adaptive DPD and PAs.

This new indirect learning architecture lends itself to efficient, yet flexible implementation for 5G NR. The forward paths, which are always functional, can be implemented in hardware (e.g. ASIC or FPGA), exploiting parallel processing. The feedback paths, which need to be functional during the weight updating phase only, can be implemented by a general purpose processor-based coprocessor. It is critical to reduce the complexity of the forward paths, especially the number of complex multipliers, to enable ultra-low latency communications in 5G NR.

The proposed scheme exploits the enormous DoFs in a massive MIMO system to reduce the number of basis functions and to mitigate its effects of crosstalk. A cascaded FIR filtering scheme, named successive refinement filtering, is proposed because it provides fast convergence to their steady-state values \cite{prandoni1998fir,yu2003lossless}.
Algorithm 1 depicts the algorithm flow of successive refinement filtering.
 \begin{algorithm}[H]
 \caption{Successive Refinement Filtering}
  \begin{algorithmic}[1]
 \renewcommand{\algorithmicrequire}{\textbf{Input:}}
 \renewcommand{\algorithmicensure}{\textbf{Output:}}
 \REQUIRE  channel matrix $\mathbf{H}$, modulated symbol $\mathbf{s}$
 \ENSURE  predistortion matrix $\mathbf{G}$, precoding matrix $\mathbf{P}$
 \\ Initialization: $ \mathbf{P} = \mathbf{P}_0$
 \WHILE{error $\geq$ threshold $\epsilon$}
    \STATE{$\mathbf{G}(m) = \text{LMS}(\mathbf{z}'(m),\mathbf{y}(m))$;}
	\STATE{DPD update: $\text{DPD} \leftarrow  \mathbf{G}(m)$;}
	\STATE{$\mathbf{P}(m) = \text{LMS}(\mathbf{r}(m),\mathbf{x}(m))$;}
	\STATE{ZFPC update: $\text{ZFPC} \leftarrow  \mathbf{P}(m)$;}
	\STATE{$m\leftarrow m+1$;}
  \ENDWHILE
 \RETURN $\mathbf{G}$ and $\mathbf{P}$
 \end{algorithmic} \label{SRF}
 \end{algorithm}
As shown in Fig. \ref{fig_block_DPD} (right), $\mathbf{y}(m)$ represents the input signal of a PA for the $m$th iteration, $\mathbf{r}(m)$ the input of the ZFPC P, $\mathbf{z}'(m)$ the input of the adaptive DPD, and $\mathbf{x}(m)$ the output of the precoding in the forward path of the proposed structure.
 
In order to give a quantitative measure of complexity of the transmitter hardware, we evaluate the floating point operations (FLOPs) \cite{afsardoost2012digital} required for the architecture of Fig. $\ref{fig_block_DPD}$ (right). The complex filtering operation requires six FLOPs per filter tap, four real multiplications and two summations. For a given delay tap $q$, the DPD output for a $(K+1)$-order Volterra series can be represent as
\begin{align}
&y_q^{K+1}(n)=\sum_{k=1}^{K+1}\omega_{k,q}|{x(n-q)}|^{k-1}x(n-q)\\\nonumber
&=\omega_{1,q}{x(n-q)}+|{x(n-q)}|\underbrace{\sum_{k=1}^K\omega_{{k+1},q}|{x(n-q)}|^{k-1}x(n-q)}_{\tilde{y}_q^K (n)}.
\end{align}
There is one complex-real multiplication for $|x(n-q)| \tilde{y}_q^K(n)$, one complex multiplication for $\omega_{1,q} x(n-q)$ (can be pre-calculated), and one complex summation\footnote{The computational complexities of $\tilde{y}_q^K (n)$ and $y_q^K (n)$ are the same.}. Therefore, the total savings of the proposed DPD when comparing it with the conventional DPD is $4(K_C-K_P )QN_t$ FLOPs, where $K_C$ denotes the order of the conventional independent DPD scheme, $K_P$ the order of the proposed scheme, $Q$ the memory depth, and $N_t$ the total number of transmit antennas in the downlink. Notice that $N_t$ is usually very large for massive MIMO configurations and hence considerable FLOPs are saved.

\section{Simulation and Experimental Results}
\subsection{Simulation Results}

\begin{figure}[h]
\centering
\includegraphics[width=3.8in]{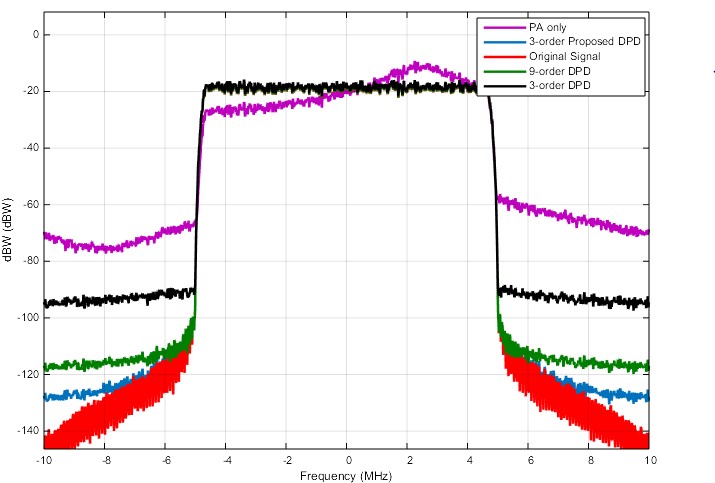}
\caption{Predistortion linearization performance in terms of spectral regrowth suppression. The PA output PSD is shown for different cases with crosstalk set to $-20$ dB for adjacent antennas. Also shown is the original signal without crosstalk.}
\label{fig_MMSE_DPD}
\end{figure}

Simulations are performed with the polynomial PA model containing memory effects. A $100 \times 10$ massive MIMO configuration with 100 transmit antenna elements is used to evaluate the proposed DPD algorithm. The simulation setup consists of RF sources that represent the transmitting paths and accurately capture the effects of crosstalk on the performance of the massive MIMO transmitter. The crosstalk effect is simulated by coupling the signal in each RF path to its adjacent paths with $20$ dB attenuation. The PA input and output are assumed to obey the Saleh model with parameters $\alpha_a=2$, $\beta_a=2.2$, $\alpha_{\phi}=2$, and $\beta_{\phi}=1$ \cite{o2009new}. The center frequency is set to 3.5 GHz and the baseband bandwidth to 10 MHz. Fig. \ref{fig_MMSE_DPD} shows the PSD output of the PA for the conventional polynomial predistorter with polynomial orders $K=3$ and $K=9$ along with the proposed polynomial predistorter with polynomial order $K=3$.

The highest trace (purple) shows the spectral regrowth of the PA output without predistortion and is seen to be approximately $-50$ dB below the in-band power level. The black trace shows the output spectrum with the conventional memory polynomial model with nonlinear degree $K=3$  and memory depth $Q=5$. It shows an out-of-band emission of $-70$ dB with respect to the in-band signal power level. The green trace is for the conventional memory polynomial model with nonlinear degree $K=9$  and memory depth $Q=5$, which reduces the regrowth to -90 dB relative to the in-band power level. The blue trace shows the output spectrum for the proposed DPD scheme described in Algorithm \ref{SRF} with nonlinear degree $K=3$  and memory depth $Q=5$. We observe that the proposed algorithm outperforms the conventional memory polynomial solution $(K=9, Q=5)$ and reduces the out-of-band emission by another $10 $ dB.
\subsection{Hybird Experimental and Simulation Results}
\begin{figure}[h]
\centering
\includegraphics[width=3.5in]{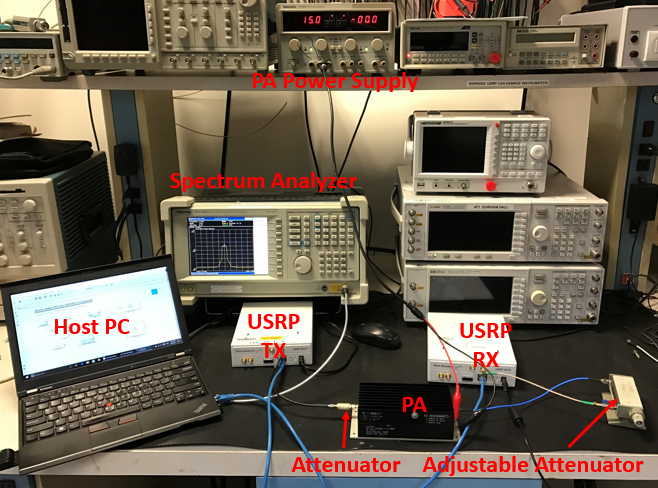}
\caption{Hybrid experimental and simulation setup (one mini-circuits ZHL-42 PA (33dB gain@3.5GHz\&15V) and two USRPs (N210) are employed).}
\label{fig_experiment}
\end{figure}

It is clear that the hardware implementation of the entire massive MIMO system requires a large number of PAs (and digital converters, mixers, etc.) since each antenna element needs an independent PA. In this subsection, we therefore introduce a hybrid experimental and simulation setup which exploits both the real and theoretical PA models to examine the proposed DPD performance with limited hardware resources (Fig. \ref{fig_experiment}). 

Our testbed features a real PA that serves as one of the 100 transmit antenna elements, whereas the others are emulated in the host PC using a theoretical PA model. Synchronization is necessary between the real path and virtual paths. The proposed hybird experimental and simulation setup consists of a mini-circuits ZHL-42 PA (33dB gain@3.5GHz\&15V) and two Ettus USRPs (model N210) which are equipped with SBX daughterboards that support radio frequencies from 400 MHz up to 4.4 GHz. A real-time spectrum analyzer (RSA3408A) is used to show the output spectrum of the ZHL-42 in the experiment. MATLAB Simulink is connected to the USRPs to run Algorithm \ref{SRF} and emulates the Saleh model for the other PAs. The USRPs are instantiated in Simulink. As shown in Fig. \ref{fig_experiment}, one transmit signal is sent to a USRP via Ethernet and from the USRP to the ZHL-42 PA after the 3.5 GHz upconversion. The amplified signal is attenuated and sent to another USRP and its output is sent to the feedback path in the host PC via an Ethernet-PCIe card after the downconversion. The rest of the transmit signals are processed by the Saleh model and go to the internal feedback path directly. The normalized mean square error (NMSE) between the output of the ZHL-42 and the undistorted input signal, for the conventional and the proposed DPD, are calculated and presented in Table \ref{tab_nmse}. It also shows the complexity comparison in terms of the number of FLOPS.
\begin{table}[H]
\def\arraystretch{1.6}\tabcolsep=5pt  
\caption{Performance comparison of no DPD, conventional DPD and the proposed DPD ($N_t=100$ and $Q=5$).}
\begin{center}
    \begin{tabular}{|l|c|c|}
    \hline
     Scheme & NMSE & Complexity \\\hline\hline
    ZHL-42 only (No DPD) & -18.41 dB & N/A\\\hline
    Conventional 3rd order DPD  & -27.38 dB & $7\times10^3$ FLOPS\\\hline
    Conventional 9th order DPD & -40.17 dB & $1.9\times 10^4$ FLOPS\\\hline
    Conventional 11th order DPD & -45.30 dB & $2.3\times 10^4$ FLOPS\\\hline
    Proposed 3rd order DPD& -45.89 dB &$7\times10^3$ FLOPS\\\hline
    \end{tabular}
\label{tab_nmse} 
\end{center}
\end{table}
The proposed 3rd order DPD scheme achieves gains of 18.5, 5.7 and 0.6 dB over the conventional 3rd order, 9th order and 11th order DPD, respectively. Note that the proposed 3rd order DPD scheme needs only 30\% of the computational power needed by the conventional 11th order DPD scheme to achieve the equivalent NMSE performance.
\section{Conclusion}
This paper has introduced a low-complexity predistorter for nonlinear PAs with memory. By exploiting the excess DoFs, our approach outperforms the conventional approach in terms of accuracy and complexity. Both simulations and hardware experiments confirm the proposed DPD model can linearize PAs better than the conventional approach. The proposed solution achieves high performance at low complexity, making it practical for 5G NR massive MIMO system deployments.

\bibliographystyle{ieeetr}
\bibliography{myreference}

\end{document}